\documentclass[hyper]{JHEP} 

\usepackage{epsfig}

\newcommand\fverb{\setbox\pippobox=\hbox\bgroup\verb}
\newcommand\fverbdo{\egroup\medskip\noindent%
			\fbox{\unhbox\pippobox}\ }
\newcommand\fverbit{\egroup\item[\fbox{\unhbox\pippobox}]}
\newbox\pippobox
\title{ Remark About dS/CFT Correspondence}
\author{by J. Kluso\v{n}\\
\footnote{On leave from Masaryk University, Brno}\\
Institute of Theoretical Physics, University of Stockholm, SCFAB\\
SE- 106 91 Stockholm, Sweden \\
and \\
Institutionen f\"or teoretisk fysik\\
BOX 803, SE- 751 08 
Uppsala, Sweden \\
E-mail: \email{josef.kluson@teorfys.uu.se}}

\preprint{\hepth{0302008}}

\abstract{In this paper we will study 
some aspects of dS/CFT correspondence. We will
focus on the relation between Witten's non-standard
de Sitter inner product and 
 correlators in the holographic dual 
 conformal field theory. We will
argue that from the definition
of Witten's inner product and conjecture 
that the Hilbert space of initial states of massive
scalar field on $\mathcal{I}^-$ in de Sitter space
corresponds to the Hilbert 
space of  states of Euclidean CFT on   
$\mathcal{I}^-$, we can obtain CFT  correlators in
 any vacuum state.}


\def\bra #1{\left<#1\right|}
\def\ket #1{\left|#1\right>}

\begin{document}
\section{Introduction}\label{first}
 Remarkable series of recent observations 
suggest that we live in accelerating space-time
\cite{Balbi:2000tg,Stompor:2001xf,
Perlmutter:1996ds,Riess:1998cb,Primack:2002th}.
However understanding the quantum theory of gravity in
de Sitter space remains one of the most important
problems in theoretical physics. 
\footnote{Very exciting and  radical proposals considering
string theory in de Sitter space  can be found
in 
\cite{Banks:zj,Banks:2001yp,Banks:2000fe}.
Alternative point of view on the
quantum gravity in three dimensional
 de Sitter space  can be found in
\cite{Smolin:2002sz}.
 For recent 
discussion of some issues of quantum theory in de
Sitter space,  see 
\cite{Banks:2002wr,Goheer:2002vf,Danielsson:2002td}.}
A correspondence relating de Sitter (dS) gravity to
the conformal field theory (CFT) theory dual was presented in 
\cite{Strominger:2001pn} and further studied in
\cite{Lin:1999gf,
Park:1998yw,Park:2001zn,Klemm:2001ea,
McInnes:2001dq,Balasubramanian:2001nb,
CarneirodaCunha:2001nw,Ghezelbash:2002vz,
Ghezelbash:2001vs,Cai:2001sn,Nojiri:2001tw,
Nojiri:2001mf,Nojiri:2001qq,Cvetic:2001bk,
Hawking:2000da,
Bousso:2000md,Balasubramanian:2001rb,Li:2001ky,Halyo:2001px,
Hull:2001ii,Danielsson:2001wt,Ogushi:2001ij}.
For further details considering this proposal, see
\cite{Spradlin:2001nb,Bousso:2001mw,Leblond:2002tf,Parikh:2002py,
Strominger:2001gp,Tolley:2001gg,
Balasubramanian:2002zh,Klemm:2002ir,Larsen:2002et,Leblond:2002ns,
Das:2002he,Kabat:2002hj}.

The dS/CFT correspondence is mainly modeled with analogy
AdS/CFT correspondence \cite{Maldacena:1997re,Gubser:1998bc,
Witten:1998qj}: Like the AdS case, the symmetries of de
Sitter space suggest that the dual theory is conformly
invariant. Mainly, it was proposed the connection 
between gravity in de Sitter space and the correlation
functions in the dual Euclidean CFT.
It is also clear that  the nature of de Sitter space 
is different from its AdS counterpart. For example, the conformal
boundaries of dS are hyper-surfaces of Euclidean signature. Further, in de
Sitter space, there are two such hyper-surfaces, the future
boundary $\mathcal{I}^+$ and the past boundary $\mathcal{I}^-$.
There was great debate in the past few months
if the proposed duality involves
single CFT \cite{Strominger:2001pn} living on
either  $\mathcal{I}^+$ or $\mathcal{I}^-$, or 
 if we should rather consider both of them as
independent parts of one single theory and study nontrivial correlation
between them, as  was suggested in very interesting paper
\cite{Balasubramanian:2002zh}. This question
was addressed in 
\cite{Leblond:2002ns} where very impressive arguments
were given that support the idea of single dual CFT.
In this paper we will consider the CFT living on
$\mathcal{I}^-$ as the fundamental one, with agreement
with \cite{Strominger:2001pn}. We will also sometimes
consider 
operators in the second CFT living on $\mathcal{I}^+$ however
as will be clear from the context the
  operators in CFT on $\mathcal{I}^+$
are determined by operators of CFT on $\mathcal{I}^-$.
 On the other
hand it is possible that the  nontrivial relation between CFT living
on $\mathcal{I}^+,\mathcal{I}^-$ could capture dynamic
of bulk quantum gravity.

However, the most striking difference from the AdS/CFT
duality is the fact that we have no rigorous
realization of the dS/CFT correspondence.
In  \cite{Strominger:2001pn,
Spradlin:2001nb}  dS/CFT duality
was proposed on the direct analogy with 
AdS/CFT correspondence when the correlation functions
in the dual CFT theory are determined through the bulk
S-matrix elements of massive scalar field.
However  it is important to stress
that there is no rigorous definition of such a duality
as opposite with Ads/CFT correspondence.  

In \cite{Witten:2001kn} different
point of view on the problem of quantum gravity
in dS space and its possible connection to the
dual CFT was suggested. It was stressed by author
that his approach   seems to give 
increasingly less information as the cosmological
constant is increased. More precisely, Witten
argued that for negative cosmological constant
the sort of reasoning, that he presented there,
  gives the boundary 
conformal field theory which, according to
\cite{Witten:2001kn} might be regarded as supplying
a dynamical principle. For zero cosmological constant
we get S-matrix. For positive cosmological constant
all what we get is definition of the Hilbert space
$\mathcal{H}_i$ of initial states at $\mathcal{I}^-$ or
equivalently  the Hilbert space
$\mathcal{H}_f$ of final states at $\mathcal{I}^+$.
In \cite{Witten:2001kn} also   definition of the
Hilbert spaces  $\mathcal{H}_{i,f}$ and 
inner product was suggested. The modified
inner product involves path integral evolution from
$\mathcal{I}^-$ to $\mathcal{I}^+$ together
with $\mathcal{CPT}$ conjugation. The concrete realization
of this proposal was given in \cite{Bousso:2001mw} in 
the case of the free massive scalar field in
$dS_3$. It was argued that the 
inner product for the bulk scalar field constructed along
the lines given in \cite{Witten:2001kn} leads to 
the  adjoint in the dual CFT  that gives
standard inner product of the boundary field theory.

In this paper we would like to apply these ideas and
argue that from the proposed modified inner product and
the relation between CFT living on
$\mathcal{I}^-$ and the bulk scalar field
 we can extract all correlation functions
in the dual CFT without any explicit form of
the conjectured dS/CFT duality. 
 We begin
with the review of the construction
of the inner product in the space of initial
states of  the free massive field
in $dS_3$ as was given in \cite{Bousso:2001mw} that
leads to the definition of the  Hilbert space $\mathcal{H}_i$ 
 of the initial states of the scalar field on
$\mathcal{I}^-$. According to the dS/CFT duality
states in $\mathcal{H}_i$ should correspond to 
the states of 
Euclidean   CFT living on $\mathcal{I}_-$. Since the construction of
the inner product presented in \cite{Bousso:2001mw} is
given in terms of $\ket{in}$ vacua with the
natural particle interpretation,  it is reasonable
to presume an existence of  
dual CFT operator $\mathcal{O}^{in}_+$
that creates these quanta from the vacuum state
$\ket{in}$. From  the same  reason we can
deduce an  existence of lowering operator
$\mathcal{O}^{in}_-$ and also that these operators
must obey canonical  commutation relations. From
 these commutation relations and definition of the inner
product we will able to reproduce all two point functions
in $\ket{in}$ vacuum state
as were given in \cite{Bousso:2001mw,Spradlin:2001nb}
as well as the correlation functions for operators defined on
$\mathcal{I}^+$ and $\mathcal{I}^-$, even it is clear that
in case of the free massive scalar field in the bulk the 
CFT operators on $\mathcal{I}^+$ are uniquely functions
of the CFT operators on $\mathcal{I}^-$. 
It is also  well known that there is one parameter family
of vacuum states 
\footnote{The quantum field theory of massive scalar field
in de Sitter space time was extensively studied in
the past, see for example 
\cite{Chernikov:zm,Tagirov:1972vv,Allen:ux,Mottola:ar,Kirsten:ug}.
For general introduction to the quantum field theory in
curved space time, see \cite{Birrell:ix,Wald:yp,Spradlin:2001pw}.
There was great discussion about consistency of general vacuum
states $\ket{\gamma}$ and their possible meaning in
cosmology, see for example
 \cite{Kaloper:2002cs,Martin:2000xs,Easther:2001fz,
Kaloper:2002uj,Brandenberger:2002hs,Banks:2002nv,Einhorn:2002nu,
Maldacena:2002vr,Danielsson:2002qh,Bergstrom:2002yd,
Danielsson:2002kx,Danielsson:2002mb,Shankaranarayanan:2002ax}.}
 for free massive scalar field in dS.
It is then natural to presume that these vacuum
states should have their description in
CFT as well. In fact, CFT correlators
in general vacuum states $\ket{\gamma}$ were
previously calculated in \cite{Bousso:2001mw,Spradlin:2001nb}.
We extend our analysis to general vacuum states as
well and we will show that we are able to
reproduce results given in 
\cite{Bousso:2001mw,Spradlin:2001nb}.  

We hope that our calculation could be
helpful for better understanding of the proposed 
dS/CFT correspondence. 
The main point of our calculations is
the conjecture  that the dual CFT 
gives an
 exact description
of the Hilbert space on $\mathcal{I}^-$.
This approach is mainly motivated by
 Witten's arguments that
dual CFT theory can only give an information about
the structure of the Hilbert space of quantum gravity of initial
states on $\mathcal{I}^-$ rather to supplies for dynamical
principle as in the case of AdS/CFT correspondence. 
It is possible that this dynamical  principle is
somewhat hidden  in the relation
between CFT operators  on $\mathcal{I}^-$ and
the operators living on $\mathcal{I}^+$ and
that this relation could support the dynamical evolution of the
bulk quantum gravity. 

It seems
to us   that the possibility to extract 
CFT correlation functions directly from the
definition of the inner product without 
any explicit realization of the dS/CFT correspondence
 could be helpful  for further study of the bulk quantum theory.
For example, there is 
Schr\"odinger picture description of the quantum field
in dS space (See, for example \cite{Long:1996wf}  and
reference there.)
 and 
at present it is not complete clear to us how this
description could have its holographic description
in dual CFT.
We hope that our approach could be useful for
addressing  this issue.

The organization of this paper is  follows. In  section
\ref{second} we review the quantisation of the 
massive scalar field in dS space. 
In  section \ref{third} we review the construction
of the inner product for the bulk scalar field as was given in
\cite{Bousso:2001mw} and we obtain all two point functions
in the dual CFT in $\ket{in}$ vacuum state. In section \ref{fourth} we will
discuss these correlators in general vacuum
states $\ket{\gamma}$. 
In conclusion 
\ref{fifth}
we summarize our results and suggest further direction
of research.
\section{Massive scalar field
in de Sitter space}\label{second}
In this section we review the quantisation
of the massive scalar field in de Sitter space
 following mainly very nice analysis
given in \cite{Bousso:2001mw}.

d-dimensional de Sitter space ($dS_d$) is described by
hyperboloid in $d+1$ dimensional Minkowski space
\footnote{With the metric $\eta_{ab}=
\mathrm{diag}(-1,1,\dots, 1)$.}
\begin{equation}\label{dS}
P(X,X)=1 \ , P(X,Y)=\eta_{ab}X^aX^b \ , 
a,b=0,\dots,d \ .
\end{equation}
In convention \cite{Bousso:2001mw} lower case
$x$ denote $d$-dimensional coordinate on
$dS_d$ and upper case $X$ to denote the corresponding
d+1 dimensional coordinate in the embedding space. 

There are many  coordinate systems that are
useful for description of $dS_d$, for recent review, see
\cite{Spradlin:2001pw,Kim:2002uz}. In this paper
we will work in the global coordinates $(\tau,\Omega)$ in
which de Sitter  metric is
\begin{equation}\label{metric}
ds^2=-d\tau^2+\cosh^2 \tau d\Omega^2_{d-1}
 \ ,
\end{equation}
where $d\Omega_{d-1}^2$ is the standard round
metric on $S^{d-1}$ . The equation of motion
of the scalar field is
\begin{equation}\label{eqgen}
(\nabla^2-m^2)\phi=0 \ ,
\nabla^2\phi=\frac{1}{\sqrt{-g}}
\partial_{\mu}(\sqrt{-g}
g^{\mu\nu}\partial_{\nu}\phi) \ 
\end{equation}
that in the metric  (\ref{metric}) has the form
\begin{equation}\label{eqglob}
-\frac{1}{\cosh^{d-1}\tau}
\partial_\tau(\cosh^{d-1}\tau \partial_{\tau}\phi)+
\frac{1}{\cosh^{d}\tau}\nabla_{S^{d-1}}^2\phi
-m^2\phi=0 \ .
\end{equation}
This equation is separable with the solution
\begin{equation}
\phi=y_L(\tau)Y_{Lj}(\Omega)
\ ,
\end{equation}
where $Y_{Lj}$ are spherical harmonic on $S^{d-1}$ obeying
\begin{equation}
\nabla^2_{S^{d-1}}Y_{Lj}=-L(L+d-2)
Y_{Lj} \ ,
\end{equation}
where $L$ is non-negative integer and $j$ is a collective
index $(j_1,\dots,j_{d-2})$.  Following
\cite{Bousso:2001mw} we choose such   $Y_{Lj}$'s 
that obey 
\begin{equation}\label{realcon}
Y_{Lj}(\Omega_A)=Y^*_{Lj}(\Omega)=
(-1)^LY_{Lj}(\Omega) \ ,
\end{equation}
where $\Omega_A$ denotes the point on $S^{d-1}$
antipodal to $\Omega$. The functions 
$Y_{Lj}$ are orthonormal 
\begin{equation}\label{ortho}
\int d\Omega Y_{Lj}(\Omega)Y^*_{L'j'}(\Omega)=
\delta_{LL'}\delta_{jj'} \ ,
\end{equation}
and complete
\begin{equation}\label{complete}
\sum_{Lj}Y_{Lj}(\Omega)Y^*_{Lj}(\Omega')=
\delta^{d-1}(\Omega,\Omega') \ .
\end{equation}
For real positive $\mu \ , 2m>(d-1)$ we find
\footnote{$\mu=\sqrt{m^2-\frac{(d-1)^2}{4}}$. 
In this paper we will consider $\mu$ real
positive only. For aspect of the quantisation of the
scalar fields with other values of $\mu$, see 
recent papers  \cite{Leblond:2002tf,Tolley:2001gg}
and reference therein.}
\begin{equation}
y^{in}_L=\frac{2^{L+d/2-1}}{\sqrt{\mu}}
\cosh^L\tau e^{(L+\frac{d-1}{2}-i\mu)\tau}
F(L+\frac{d-1}{2},L+\frac{d-1}{2}-i\mu \ ,
1-i\mu, -e^{2\tau}) 
\end{equation}
which together with 
its complex conjugate are two independent solutions.
The normalization is fixed that these modes are orthonormal
with respect to the standard inner product 
\begin{equation}\label{KGinnerprod}
(\phi_{Lj},\phi_{L'j'})=-i\int d\Omega \cosh^{d-1}\tau 
\left(\partial_{\tau}\phi^*_{Lj}\phi_{L'j'}-
\phi^*_{Lj}\partial_{\tau}\phi_{L'j'}\right)=
\delta_{LL'}\delta_{jj'}
\ ,
\end{equation}
where the integral is performed over $S^{d-1}$ at fixed
 $\tau$. As is well known (See, for example
\cite{Birrell:ix}.) this inner  product is independent
on  time slicing  for modes obeying (\ref{eqgen})
so that we can evaluate it in the
limit  $\tau\rightarrow
-\infty$ when 
 $F\rightarrow 1  \ , \cosh \tau\rightarrow
\frac{1}{2}e^{-\tau}$ and consequently
\begin{equation}\label{inlim}
\lim_{\tau\rightarrow -\infty}\phi^{in}_{Lj}
(x)=
\frac{2^{d/2-1}}{\sqrt{\mu}}
e^{(\frac{d-1}{2}-i\mu)\tau}Y_{Lj}(\Omega) \ .
\end{equation}
Then it is  simple task to show validity (\ref{KGinnerprod}). 

From (\ref{inlim}) it is clear
 that  $\phi^{in}$  modes correspond
to the positive frequency modes with respect to 
the global time $\tau$ near the asymptotic past
and hence they represent incoming particle states.
Then we define $\ket{in}$ vacuum as a state 
which is annihilated by the lowering operator associated
to $\phi^{in}$. Physically, $\ket{in}$ is the state
with no incoming particles on $\mathcal{I}^-$. 

We also see that (\ref{eqglob}) is
invariant under time reversal
 $\tau \rightarrow -\tau$. Hence 
we can obtain another pair of linearly independent solutions
by defining
\begin{equation}
y^{out}_L(\tau)=y^{in *}_L(-\tau) \ 
\end{equation}
with the asymptotic behavior
\begin{equation}
\lim_{\tau\rightarrow
\infty} y^{out}\rightarrow \frac{2^{d/2-1}}{\sqrt{\mu}}
e^{-(\frac{d-1}{2}+i\mu)\tau} 
\end{equation}
which implies  that  modes
\begin{equation}
\phi^{out}_{Lj}(x)=
y^{out}_L(\tau)Y_{Lj}(\Omega) \ 
\end{equation}
are positive frequency modes with respect to the
global time $\tau$ near the asymptotic  future
and  
 represent outgoing particle states.
As in the case of  $\phi^{in}_{Lj}$ we
can define  vacuum state $\ket{out}$  which is  the state
 annihilated by
the lowering operators associated to 
$\phi^{out}$. 
Generally $\ket{in}$ state is not the
same as $\ket{out}$ state however as was recently reviewed in
\cite{Bousso:2001mw}, for odd dimensional
de Sitter space we can identify them 
\begin{equation}
\ket{in}=\ket{out}  
\mathrm{in \  odd \  dimensions} \ .
\end{equation}
It follows that
 there is no particle production in the
sense that if no particles are coming in
from $\mathcal{I}^-$ no particles will
go out on $\mathcal{I}^+$. This is in contrast
to the even-dimensional case for which
there is always some particle production.

Note that (\ref{inlim})  implies 
asymptotic behavior of $\phi^{in}$ near
$\mathcal{I}^-$
\begin{equation}
\phi^{in}_{\pm}\sim e^{h_{\pm}\tau} \ ,
h_{\pm}=\frac{d-1}{2}\pm i\mu \ ,
\mathrm{for}  \  \tau \  \rightarrow -\infty  \ .
\end{equation}
As was conjectured in \cite{Strominger:2001pn} 
 $\phi^{in}_-$ is  dual to
the operators of weight $h^+$ in the CFT living
on $\mathcal{I}^-$.
 In the same way 
$\phi^{in }_+$ is  dual to operator
of conformal weight $h_-$. 

As was shown in \cite{Bousso:2001mw} it is useful
for  the definition of general vacuum states 
$\ket{\gamma}$ to introduce  rescaled global
modes
\begin{eqnarray}\label{globrescal}
\tilde{\phi}^{in}_{Lj}(x)
=e^{i\theta_L}y^{in}_{L}(\tau)
Y_{Lj}(\Omega) \ , \nonumber \\
\tilde{\phi}^{out}_{Lj}(x)
=e^{-i\theta_L}y^{out}_{L}(\tau)
Y_{Lj}(\Omega) 
\ , \nonumber \\
\end{eqnarray}
where the phase $e^{i\theta_L}$ is defined
\begin{equation}
e^{-2i\theta_L}=(-1)^{L-\frac{d-1}{2}}
\frac{\Gamma(-i\mu)\Gamma\left(L+\frac{d-1}{2}
+i\mu\right)}
{\Gamma(i\mu)\Gamma\left(L+\frac{d-1}{2}-i\mu\right)
} \ .
\end{equation}
These rescaled modes will be used in the
discussion of the general vacuum state in
section \ref{fourth}.

In the following we will also need to
know how  at
the general point in the bulk
 the  field  is determined from its
value on $\mathcal{I}^-$. It can be
shown that these fields are related through
\begin{eqnarray}\label{bulkboundary}
\phi(x)=i\int_{\mathcal{I}_-}d\Omega' 
\sqrt{g(\tau')}\left(\partial_{\tau'}
G_C(x,x')\phi(x')-G_C(x,x')\partial_{\tau'}\phi(x')
\right) \ ,
\nonumber \\
G_C(x,x')=G_E(x,x')-G_E(x',x) \ ,
G_E(x,x')=\bra{E}\phi(x)\phi(x')\ket{E} \ , \nonumber \\
\end{eqnarray}
where $\ket{E}$ is Euclidean vacuum and $G_E(x,x')$ is
Wightman function characterized given vacuum state, for more
details and very nice discussion, see again \cite{Bousso:2001mw}.
It is important to stress that $G_c$ does not
depend on vacuum state that defines Wightmann function.
For our purposes it is important to find limiting value of
$G_C(x,x')$ for $\tau\rightarrow \infty$. It can be shown that 
\cite{Bousso:2001mw} 
\begin{eqnarray}
\lim_{\tau'\rightarrow -\infty \ ,
\tau \rightarrow \infty}
G_E(x,x')=-e^{h_+(\tau'-\tau)}e^{-\pi \mu}
\triangle_+(\Omega',\Omega_A)-
e^{h_-(\tau'-\tau)}e^{\pi\mu}
\triangle_-(\Omega',\Omega_A) \ , \nonumber \\
\lim_{\tau'\rightarrow -\infty \ ,
\tau \rightarrow \infty}
G_E(x',x)=-e^{h_+(\tau'-\tau)}e^{\pi \mu}
\triangle_+(\Omega',\Omega_A)-
e^{h_-(\tau'-\tau)}e^{-\pi\mu}
\triangle_-(\Omega',\Omega_A) \ , \nonumber \\
\end{eqnarray}
where
\begin{equation}\label{triangle}
\triangle_{\pm}
(\Omega,\Omega')=
-\frac{1}{\mu\sinh\pi\mu}\sum_{Lj}
e^{\mp 2i\theta_L}Y_{Lj}(\Omega)Y_{Lj}(\Omega') \ 
\end{equation}
are two point functions for a conformal field
of dimension $h_{\pm}$ on the sphere.
Then we get
\begin{equation}
\lim_{\tau'\rightarrow -\infty \ ,
\tau \rightarrow \infty}
G_C(x,x')=2e^{h_+(\tau'-\tau)}
\sinh \pi\mu \triangle_+(\Omega',\Omega_A)
-2\sinh \pi\mu e^{h_-(\tau'-\tau)}
\triangle_-(\Omega',\Omega_A) \ .
\end{equation}
Let us introduce $\phi^{in}_{\pm}(\Omega)\ ,
\phi^{out}_{\pm}(\Omega)$ fields on
$\mathcal{I}^- \ , \mathcal{I}^+$ by
\begin{eqnarray}\label{philimit}
\lim_{\tau \rightarrow -\infty}
\phi(\tau,\Omega)=\phi^{in}_+(\Omega)e^{h_+\tau}+
\phi^{in}_-(\Omega)\phi^{h_-\tau} \ , \nonumber \\
\lim_{\tau \rightarrow \infty}
\phi(\tau,\Omega_A)=\phi^{out}_+(\Omega)e^{h_+\tau}+
\phi^{out}_-(\Omega)\phi^{h_-\tau} \ , \nonumber \\
\end{eqnarray}
where, according to \cite{Strominger:2001pn,
Bousso:2001mw} $\phi^{out}_{\pm}$ have been defined
with an antipodal inversion relative to 
$\phi^{in}_{\pm}$. Asymptotic form 
(\ref{inlim}) for $d=3$ implies 
\begin{eqnarray}
\phi^{in}_+(\Omega)=
\sqrt{\frac{2}{\mu}}\sum_{Lj}
b^{in}_{Lj}Y^*_{Lj}(\Omega)
 \ , \phi^{in}_-(\Omega)=
\sqrt{\frac{2}{\mu}}\sum_{Lj}
a^{in}_{Lj}Y_{Lj}(\Omega)
\nonumber \\
\phi^{out}_+(\Omega)=
\sqrt{\frac{2}{\mu}}\sum_{Lj}
a^{out}_{Lj}Y_{Lj}(\Omega_A)
 \ , \phi^{out}_-(\Omega)=
\sqrt{\frac{2}{\mu}}\sum_{Lj}
b^{out}_{Lj}Y^*_{Lj}(\Omega_A)
 \ .\nonumber \\
\end{eqnarray}
Then (\ref{bulkboundary}) gives
\begin{equation}\label{outin}
\phi^{out}_{\pm}(\Omega)=-\mu\sinh\pi\mu
\int d\Omega \triangle_{\pm}(\Omega,
\Omega')\phi^{in}_{\mp}(\Omega') \ .
\end{equation}
In the next section we
use many of the facts reviewed above 
 for the  construction of the non-standard
inner product 
\cite{Witten:2001kn} and its  realization
in the case of free massive scalar field 
in \cite{Bousso:2001mw}. 
\section{CFT correlators from 
inner product}\label{third}
In \cite{Witten:2001kn}  
Witten suggested how
to construct  Hilbert spaces of
initial and final states of quantum gravity
in de Sitter space, together with
the definition of inner product. 
This definition is based 
on the  path integral evolution from $\mathcal{I}^-$ to
$\mathcal{I}^+$ together with $\mathcal{CPT}$
conjugation. Since we believe that there
is an exact description of the quantum gravity
and quantum field theory  
in  de Sitter space in terms of dual 
CFT living  on $\mathcal{I}^-$ we expect
that the Hilbert space of the initial
states of scalar field in the bulk  is described
using CFT operators acting on some vacuum state.
We will show that from the definition of the inner
product  according to Witten and its very nice analysis
in \cite{Bousso:2001mw}   we can obtain
all correlation functions in the dual CFT. 

To begin with,
we briefly review  discrete symmetries
$\mathcal{C, \ P}$ and $\mathcal{T}$ in
de Sitter space, following \cite{Bousso:2001mw}.
As in this paper we restrict ourselves to the
three dimensional de Sitter space. 

Let us consider the real scalar field so that the
operation of charge conjugation $\mathcal{C}$ is trivial.
We define action of the following discrete symmetry operations
on this field
\begin{equation}\label{TPphi}
\mathcal{P}\phi(x)\mathcal{P}=\phi(Px) \ ,
\mathcal{T}\phi(x)\mathcal{T}=\phi(Tx) \ ,
\end{equation}
where we have two discrete symmetries of de Sitter
space $P$ and $T$ that are defined as
  \cite{Bousso:2001mw}
\begin{eqnarray}
PX^0=X^0 \ , PX^1=X^1 \ ,
PX^2=-X^2 \ , PX^3=-X^3 \ , \nonumber \\
TX^0=-X^0 \ , TX^1=X^1 \ ,
TX^2=X^2 \ , TX^3=X^3 \  \nonumber \\
\end{eqnarray}
that clearly leave (\ref{dS}) invariant.
Bulk scalar field $\phi$ 
can be written as
\begin{eqnarray}
\phi(\tau,\Omega)=
\sum_{L,j}
\left(a^{in}_{Lj}y^{in}_L(\tau)Y_{Lj}(\Omega)
+b^{in}_{Lj}y^{in*}_L(\tau)Y^*_{Lj}(\Omega)\right)
 \ , \nonumber \\
\phi(\tau,\Omega)=
\sum_{L,j}
\left(a^{out}_{Lj}y^{out}_L(\tau)Y_{Lj}(\Omega)
+b^{out}_{Lj}y^{out*}_L(\tau)Y^*_{Lj}(\Omega)\right)
\ . \nonumber \\
\end{eqnarray}
As in \cite{Bousso:2001mw} we have introduced
lowering and raising operators $a_{Lj},b_{Lj}$ respectively
and are not assuming that $a^{\dag}_{Lj}=b_{Lj}$. 
Then the action of the discrete symmetries $\mathcal{P},
\mathcal{T}$ was defined in such a way that 
their definition reproduces (\ref{TPphi}).
More precisely we have
\begin{equation}
\mathcal{P}a^{in}_{Lj}\mathcal{P}=
(-1)^ja^{in}_{Lj} \ ,
\mathcal{P}b^{in}_{Lj}\mathcal{P}= 
(-1)^jb^{in}_{Lj}  \ .
\end{equation}
Since $Y_{Lj}(P\Omega)=(-1)^jY_{Lj}(\Omega)$ we immediately
see that this definition reproduces
the first line in  (\ref{TPphi}). In the same way
\begin{equation}
\mathcal{T}a^{in}_{Lj}\mathcal{T}=
(-1)^La^{in}_{Lj} \ ,
\mathcal{T}b^{in}_{Lj}\mathcal{T}=
(-1)^Lb^{in}_{Lj}  \ .
\end{equation}
We note that time reversal transformation $\mathcal{T}$
is anti linear operator which combines unitary operator
$U$ with complex conjugation $K$ of the functions. 
Then it can be shown 
\cite{Bousso:2001mw}
\begin{eqnarray}\label{ptphi}
\mathcal{PT}\phi^{in}_+(\Omega)
\mathcal{PT}=\phi^{out}_-(P\Omega_A)
 \ , \nonumber \\
\mathcal{PT}\phi^{out}_+(\Omega)
\mathcal{PT}=\phi^{in}_-(P\Omega_A)
\ . \nonumber \\
\end{eqnarray}
Now we will review  the construction
of the Witten's inner product
\cite{Witten:2001kn} 
 given in
\cite{Bousso:2001mw}.
Let us consider one particle  states on $\mathcal{I}^{\pm}$ 
\begin{eqnarray}
\ket{\Psi^{in}}=
\int d\Omega \Psi^{in}(\Omega)\phi^{in}_+(\Omega)\ket{in} \ ,
\nonumber \\
\ket{\Psi^{out}}=
\int d\Omega \Psi^{out}(\Omega)\phi^{out}_-(\Omega)\ket{out} \ .
\nonumber \\
\end{eqnarray}
Using (\ref{outin}) 
we can express  $\ket{\Psi^{out}}$ as
\begin{equation}
\ket{\Psi^{out}}=-\mu\sinh\pi\mu
\int d\Omega d\Omega'
 \Psi^{out}(\Omega)\triangle_-(\Omega,\Omega')
\phi^{in}_+(\Omega')\ket{in} \ .
\end{equation}
This  defines the bilinear pairing
\begin{equation}\label{pairing1}
\left(\Psi^{out}|\Psi^{in}\right)=-
\mu\sinh\pi\mu\int d\Omega d\Omega'
\Psi^{out}(\Omega')\triangle_-
(\Omega',\Omega)\Psi^{in}(\Omega) \ .
\end{equation}
 Now we apply the 
$\mathcal{CPT}$ conjugation to any state
on $\mathcal{I}^-$ to give a state on
$\mathcal{I}^+$ 
\begin{equation}
\mathcal{CPT}\ket{\Psi^{in}}=
\int d\Omega \Psi^{*in}(\Omega)
\phi^{out}_-(P\Omega_A)\ket{in}=
\int d\Omega \Psi^{in *}
(P\Omega_A)
\phi^{out}_-(\Omega)\ket{in} \ .
\end{equation}
Then we can use the previous pairing to define the inner
product on $\mathcal{I}^-$ 
\begin{equation}\label{QGprod}
\left<\Psi^{in}|
\Gamma^{out}\right>=-\mu\sinh\pi\mu
\int d\Omega d\Omega'
\Psi^{in *}(\Omega)\triangle_-(P\Omega_A,
\Omega')\Gamma^{in}(\Omega') \ .
\end{equation}
According to dS/CFT conjecture any  state in the
Hilbert space $\mathcal{H}_i$ should have
description in  the dual CFT. Using  well
known CFT  state-operator correspondence 
 we expect that this state  is given by action
of some   operator 
$\mathcal{O}^{in}_+$
 on the vacuum state
$\ket{in}$. Let us denote
the complete set of these operators as $\mathcal{O}^{in}_{+Lj}
$ where the quantum numbers  $L,j$ have
the same meaning as in section \ref{second}.
 Then any
state in the Hilbert space $\mathcal{H}_i$
can be written as
\begin{equation}
\ket{\Gamma}=\sum_{L,j} f_{Lj} 
\mathcal{O}^{in}_{+Lj}
\ket{in}=\sqrt{\frac{2}{\mu}}\int d\Omega 
 \Psi^{in}(\Omega)\mathcal{O}^{in}_+(\Omega)\ket{in} \ 
\end{equation}
using 
\begin{equation}
\mathcal{O}^{in}_+(\Omega)=\sqrt{\frac{\mu}{2}}
\sum_{Lj} Y_{Lj}^{in*}(\Omega)\mathcal{O}_{+Lj}
^{in} \ , \Psi^{in}(\Omega)=
\sum_{Lj} f_{Lj}
Y_{Lj}^{in}(\Omega) \ .
\end{equation}
We define  $\mathcal{O}^{in}_{+Lj}$ as creation  operator
in  CFT Hilbert space $\mathcal{H}_i$.
 It is then natural to postulate an  existence
of the annihilation  operator $\mathcal{O}^{in}_{-Lj}$
with the canonical commutation relation
\begin{equation}
[\mathcal{O}^{in}_{-Lj},
\mathcal{O}^{in}_{+L'j'}]=\delta_{LL'}\delta_{jj'} \ 
\end{equation}
which implies
\begin{eqnarray}
[\mathcal{O}^{in}_{-}(\Omega),
\mathcal{O}^{in}_{+}(\Omega')]=
\left[\sqrt{\frac{\mu}{2}}
\sum_{Lj}  Y_{Lj}^* (\Omega)\mathcal{O}^{in}_{-Lj}
,\sqrt{\frac{\mu}{2}}
 Y_{L'j'} (\Omega')\mathcal{O}^{in}_{+L'j'}
\right]
= \nonumber \\
=\frac{\mu}{2} 
\sum_{Lj,L'j'}Y_{Lj}^* (\Omega)
 Y_{L'j'} (\Omega')\delta_{LL'}
\delta_{jj'}=\frac{\mu}{2}\delta(\Omega,\Omega') \ .
\nonumber \\ 
\end{eqnarray}
From the upper expression we immediately get
 two  point function 
(Note that the commutator is pure number and hence
its vacuum expectation value  is
the same for all vacuum sates.)
\begin{eqnarray}
\bra{in}[\mathcal{O}^{in}_-
(\Omega),\mathcal{O}^{in}_+
(\Omega')]\ket{in}=
\frac{\mu}{2} \delta (\Omega,\Omega')
\Rightarrow   \bra{in}\mathcal{O}^{in}_-
(\Omega)\mathcal{O}^{in}_+
(\Omega')\ket{in}=
\frac{\mu}{2} \delta (\Omega,\Omega')
\nonumber \\
\end{eqnarray}
using
\begin{equation}
\mathcal{O}^{in}_-(\Omega)\ket{in}=0 \ .
\end{equation}
In order to determine other two point
functions we  use
 the definition of the inner product
given above. This is the main point of
our analysis that should deserve deeper explanation.
In the usual dS/CFT correspondence the
CFT correlators are determined from 
the possible equivalence of CFT partition
function and S-matrix in the bulk, which
is the proposal given in 
\cite{Strominger:2001pn,Spradlin:2001nb}.
The motivation for this proposal comes 
from the AdS/CFT correspondence. However when
we adopt  Witten's arguments
\cite{Witten:2001kn} considering the dependence of
the 
holographic description of the quantum gravity
on cosmological constant and  his conclusion
that CFT  could give information
about Hilbert space of initial states $\mathcal{H}_i$
it seems to us very natural to search for
dS/CFT correspondence in the structure
of this Hilbert space and in particular in
the definition of the inner product in $\mathcal{H}_i$. 
We will see that this analysis really allows
us to get desired two point functions.  
In the CFT the inner product is defined as
\begin{equation}\label{CFTscalar}
\left<\Psi^{in}|
\Gamma^{out}\right>=
\frac{2}{\mu}\int d\Omega d\Omega'
\Psi^{in*}(\Omega)\left<(O_{+}^{in}(\Omega))^{\dag}
\mathcal{O}_+^{in}(\Omega')\right>
\Gamma^{in}(\Omega')
\end{equation}
so that comparing this expression with
(\ref{QGprod}) gives
\begin{equation}\label{eq1}
\frac{2}{\mu}\left<(O_{+}^{in}(\Omega))^{\dag}
\mathcal{O}_+^{in}(\Omega')\right>=-\mu\sinh\pi\mu
\triangle_-(P\Omega_A,\Omega') \ .
\end{equation}
Even if we do not know the precise form
of the dS/CFT correspondence we can  at least
try to suggest the interaction between 
CFT operators and the bulk field. We 
expect that the interaction between
bulk fields and boundary CFT operators
 has following local
form
\begin{eqnarray}\label{interaction}
\int d\Omega \left(\mathcal{O}_+^{in}
(\Omega)\phi_+^{in}(\Omega)+
\mathcal{O}_-^{in}
(\Omega)\phi_-^{in}(\Omega)\right) \ , \nonumber \\
\int d\Omega \left(\mathcal{O}_+^{out}
(\Omega)\phi_+^{out}(\Omega)+
\mathcal{O}_-^{out}
(\Omega)\phi_-^{out}(\Omega)\right) \ \nonumber \\
\end{eqnarray}
where we now consider the second CFT living
on $\mathcal{I}^+$ even if it will be clear
in the moment that there is unique
relation between CFT operators $\mathcal{O}^{in},
\mathcal{O}^{out}$ at least in 
the case of the free massive field in the bulk.
Now when we apply (\ref{outin}) in
(\ref{interaction}) we obtain
\begin{eqnarray}
\int d\Omega \left(
\mathcal{O}_+^{in}
(\Omega)\phi_+^{in}(\Omega)+
\mathcal{O}_-^{in}
(\Omega)\phi_-^{in}(\Omega)\right)
=\nonumber \\
=-\mu\sinh \pi\mu\int d\Omega d\Omega'
\left(\mathcal{O}_+^{in}(\Omega)
\triangle_+(\Omega,\Omega')\phi_-^{out}(\Omega')
+\right. \nonumber \\
\left.+\mathcal{O}_-^{in}(\Omega)
\triangle_-(\Omega,\Omega')\phi_+^{out}(\Omega')
\right) \nonumber \\
\end{eqnarray}
so that when we compare this
expression  with the second line in 
(\ref{interaction}) we obtain the relation
between $in$ and $out$  CFT operators
\begin{equation}\label{CFTinout}
\mathcal{O}^{out}_{\mp}=
-\mu\sinh\pi\mu\int d\Omega'\triangle_{\pm}
(\Omega,\Omega')\mathcal{O}^{in}_{\pm}(\Omega') 
\end{equation}
with agreement with 
\cite{Spradlin:2001nb}. We see that it is natural
to define the adjoint operation in the
CFT as  
\begin{equation}\label{CFTherm}
(\mathcal{O}^{in}_{\pm}(\Omega))^{\dag}=
\mathcal{O}^{out}_{\pm}(P\Omega_A) \ ,
(\mathcal{O}^{out}_{\pm}(\Omega))^{\dag}=
\mathcal{O}^{in}_{\pm}(P\Omega_A)
\end{equation}
since now (\ref{eq1})
is equal to
\begin{eqnarray}
\frac{2}{\mu}\left<(\mathcal{O}_{+}^{in}(\Omega))^{\dag}
\mathcal{O}_+^{in}(\Omega')\right>=
\frac{2}{\mu}\left<\mathcal{O}^{out}_+(P\Omega_A)
\mathcal{O}_+^{in}(\Omega')\right>=
\nonumber \\
=-2\sinh\pi\mu\int d\Omega''\triangle_{-}
(P\Omega_A,\Omega'')
\left<\mathcal{O}^{in}_{-}(\Omega'') 
\mathcal{O}^{in}_+(\Omega')\right>=
\nonumber \\
=-2\sinh\pi\mu\int d\Omega''\triangle_-
(P\Omega_A,\Omega'')
\frac{\mu}{2}\delta (\Omega'',\Omega')=
-\mu\sinh\pi\mu
\triangle_-(P\Omega_A,\Omega')
 \nonumber \\
\end{eqnarray}
 using (\ref{CFTinout}).
From the definition of  adjoint and
the inner product we can get all correlation
functions in CFT. First of all,
since  $\mathcal{O}^{in}_-$ annihilates
vacuum state we immediately obtain  following correlators
\begin{equation}
\left<\mathcal{O}^{in}_-(\Omega)
\mathcal{O}^{in}_-(\Omega')\right>=0 \ ,
\left<\mathcal{O}^{in}_+(\Omega)
\mathcal{O}^{in}_-(\Omega')\right>=0 
\end{equation}
with agreement with \cite{Bousso:2001mw,Spradlin:2001nb}.
We also have
\begin{eqnarray}
\left<\mathcal{O}^{in}_+(\Omega)
\mathcal{O}^{in}_+(\Omega')\right>
\sim\int d\Omega''\triangle_-(\Omega,\Omega'')
\left<\mathcal{O}^{out}_-(\Omega'')
\mathcal{O}^{in}_+(\Omega')\right>\sim
\nonumber \\\sim
\left<(\mathcal{O}_-(P\Omega''_A))^{\dag}
\mathcal{O}^+(\Omega')\right>=
0 \ , \nonumber \\
\end{eqnarray}
where we  used the definition of 
adjoint given above and the fact that
if $\mathcal{O}^{in}_-$ annihilates 
ket $\ket{in}$ then  its adjoint
annihilates  bra  $\bra{in}$. 

In this section we calculated the
two point functions in the CFT defined on
 $\mathcal{I}^-$.
This calculation was based on the analysis
of the modified inner product 
\cite{Bousso:2001mw}. Since we
began  from the vacuum state $\ket{in}$  
with its  natural particle interpretation
we could  introduce
 creation and annihilation operators
in CFT  that 
 create or annihilate given
quanta. 
Since it is well known that there is one parameter
family of different vacuum states of the bulk 
massive field in dS  we would like to see 
how CFT  operators $\mathcal{O}^{in}_+,\mathcal{O}^{in}_-$
act on general vacuum state $\ket{\gamma}$.
This analysis will be done in the next section.

\section{Correlation functions  in $\gamma$ vacua}
\label{fourth}
In this section we will calculate  
CFT correlators in other vacuum states
$\ket{\gamma}$. Firstly
 we review the basic facts about
different vacuum states of the scalar
field in de Sitter space, following mainly
\cite{Bousso:2001mw}. 
Let us consider the free massive field in 
de Sitter space with the mode expansion
\begin{equation}
\phi(x)=\sum_n \left[
a_n\phi_n(x)+b_n\phi^*_n(x)\right] \ ,
\end{equation}
where subscript $n$ labels 
states of given mode $\phi_n$ that
obeys (\ref{eqgen}). In global coordinates, for example,
$n$ means $L,j$ given in  the previous section.
We have also denoted, following \cite{Bousso:2001mw}
lowering and raising operators as $a_n$ and
$b_n$ with commutation
relation $[a_n,b_m]=\delta_{nm}$. 
The Wightman function is  defined as
\begin{equation}
G_{\gamma}(x,x')=\bra{\gamma}
\phi(x)\phi(x')\ket{\gamma}=
\sum_n
\phi_n(x)
\phi^*_n(x')
\end{equation}
characterized vacuum state $\ket{\gamma}$ that
is defined to be the state annihilated
by operators associated to positive modes $\phi_n$
\begin{equation}
a_n\ket{\gamma}=0 \ .
\end{equation}
Since  modes $\phi_n$ satisfy the
de Sitter space wave function
(\ref{eqgen}), it is easy to see that Wightman function
obeys 
\begin{equation}
(\nabla^2-m^2)G_{\gamma}(x,x')=0 \ .
\end{equation}
Let $\tilde{\phi}^{in}_n(x)$ denote the positive frequency modes
(\ref{globrescal}) associated to $\ket{in}$ vacuum.
 Let us consider a new set of modes
related by the MA (Mottola-Allen) transform
\begin{eqnarray}
\hat{\phi}_n=N_{\gamma}(\tilde{\phi}
^{in}_n-e^{\gamma}\tilde{\phi}_n^{in*})
\ , N_{\gamma}\equiv\frac{1}{\sqrt{1-e^{\gamma+\gamma^*}}} \ ,
\nonumber \\
\tilde{\phi}_n=N_{\gamma}(\hat{\phi}_n+
e^{\gamma}\hat{\phi}_n^{\dag}) \ ,
\nonumber \\
\end{eqnarray}
where $\gamma$ can be any complex number with $\mathrm{Re}
\gamma<0$. These modes can be used to define new operators
$\hat{a}_n \ , \hat{b}_n$ and hence the vacuum
state $\ket{\gamma}$ that is annihilated by operators
$\hat{a}_n$
 that multiply the positive frequency
solutions $\hat{\phi}_n$ 
in the expansion of the scalar field
field as 
\begin{eqnarray}
\phi(x)=\sum_n\left[\hat{a}_n\hat{\phi}_n(x)+
\hat{b}_n\hat{\phi}_n^*(x)\right]=\nonumber \\
=\sum_n\left[\tilde{a}_n^{in}\tilde{\phi}^{in}_n+
\tilde{b_n}^{in}\tilde{\phi}_n^{in*}\right]
\Rightarrow
\nonumber \\
\Rightarrow
\hat{a}_n=N_{\gamma}(\tilde{a}_n^{in}+
e^{\gamma^*}
\tilde{b}_n^{in}) \ , 
\hat{b}_n=N_{\gamma}(\tilde{b}_n^{in}
+
e^{\gamma}
\tilde{a}^{in}_n) \ . 
\nonumber \\ 
\end{eqnarray}
This relation can be written as
\begin{equation}\label{defUK}
\hat{a}_n=\mathcal{U}\tilde{a}_n^{in}
\mathcal{U}^{-1} \ ,
\mathcal{U}=e^K \ , 
K=\sum_n  (c(\tilde{b}_n^{in})^2-\overline{c}
(\tilde{a}_n^{in})^2) \ 
\end{equation}
with
\begin{equation}
c=-\frac{1}{4}\left(
\ln\tanh(\frac{Re \gamma}{2})\right)
e^{iIm\gamma}\equiv Ae^{-iIm \gamma} \ .
\end{equation}
To prove  that $\hat{a}$ can
be written as in (\ref{defUK}) 
we use following commutation
relations
\begin{equation}
[K,\tilde{a}_n^{in}]=
\sum_mc[(\tilde{b}_m^{in})^2,\tilde{a}_n^{in}]=
-2c \tilde{b}_m^{in} \ ,
[K,[K,\tilde{a}^{in}]]=-4c\overline{c}\tilde{a}_n^{in} \ 
\end{equation} 
so that
\begin{eqnarray}
\mathcal{U}\tilde{a}_n^{in}
\mathcal{U}^{-1}=\tilde{a}_n^{in}+\sum_{k=1}^{\infty}
\frac{1}{k!}\overbrace{[[K,\tilde{a}^{in}_n],\dots]}^{k}=
(-2c+\frac{8c^2\overline{c}}{6}+\dots)\tilde{b}_n^{in}+\nonumber \\
+
(1-\frac{4c\overline{c}}{2}-\frac{16(c\overline{c})^2}{24}
+\dots)\tilde{a}_n^{in}=
\sinh (-2A)e^{-Im\gamma}\tilde{b}_n^{in}+\cosh (-2A)
\tilde{a}_n^{in} \ .
\nonumber \\
\end{eqnarray}
Consequently the vacuum state $\ket{\gamma}$
can be written as 
\begin{equation}
\ket{\gamma}=\mathcal{U}\ket{in} \ .
\end{equation}
We can also express $\mathcal{U}$ using
in and out fields as
\begin{equation}
\mathcal{U}=
\exp\left(c(\gamma)\frac{\mu}{2}\int d^2\Omega 
\phi^{in}_+(\Omega)\phi^{out}_-(\Omega)+
c(\overline{\gamma})\frac{\mu}{2}
\int d\Omega \phi^{in}_-(\Omega)
\phi^{out}_+(\Omega)\right) \  .
\end{equation}
To see this note that in the definition of the MA
transformation  we  used
rescaled  fields 
\cite{Bousso:2001mw} 
\begin{equation}
\tilde{\phi}_{Lj}^{in}(x)=e^{i\theta_L}y^{in}_L(\tau)
Y_{Lj}(\Omega) \ .
\end{equation}
This implies that the lowering and raising operators given
in the MA transformation formula are related to the
operators defined using $\phi^{in}_+(\Omega) \ ,
\phi^{in}_-(\Omega)$ as 
$\tilde{b}_n^{in}=e^{i\theta_L}b_n^{in} \ ,
\tilde{a}_n^{in}= e^{-i\theta_L}b_n^{in}$.
 Then in the 
variables corresponding to
$\phi^{in,out}_{\pm}(\Omega)$ modes
we can express  operator $K$ as
\begin{eqnarray}
K=\sum_n (c e^{2i\theta_L}(b_n^{in})^2-
\overline{c}e^{-2i\theta_L}(a_n^{in})^2)=
\nonumber \\
=c\frac{\mu}{2}
\int d\Omega 
\phi^{in}_+(\Omega)\phi^{out}_-(\Omega)-
\overline{c}\frac{\mu}{2}
\int d\Omega 
\phi^{in}_-(\Omega)\phi^{out}_+(\Omega)
\ . \nonumber \\
\end{eqnarray}
According to dS/CFT correspondence 
we can presume an existence of one
parameter family of vacuum states
$\ket{\gamma}$ in CFT as well and that
these states are given as
\begin{equation}
\ket{\gamma}=\mathcal{U}\ket{in} \ ,
\mathcal{U}=e^K \ ,
K=c(\gamma)\frac{2}{\mu}\int d\Omega
\mathcal{O}^{in}_+(\Omega)
\mathcal{O}_-^{out}(\Omega)-
\overline{c}(\gamma)\frac{2}{\mu}\int d\Omega
\mathcal{O}^{in}_-(\Omega)
\mathcal{O}_+^{out}(\Omega) \ .
\end{equation}
Using (\ref{CFTherm})we can easily show
that 
\begin{equation}
K^{\dag}=-K 
\end{equation}
with respect to the adjoint given in the previous section.
Then   two point functions of 
CFT operators in general vacuum state $\ket{\gamma}$ 
are
\begin{eqnarray}\label{twogen1}
\bra{\gamma}\mathcal{O}^{in}_{\pm}(\Omega)
\mathcal{O}^{in}_{\pm}(\Omega')\ket{\gamma}=
\left<e^{-K}\mathcal{O}_{\pm}(\Omega)
e^Ke^{-K}
\mathcal{O}^{in}_{\pm}(\Omega')e^K\right> \ ,
\nonumber \\
\bra{\gamma}\mathcal{O}^{in}_{\pm}(\Omega)
\mathcal{O}^{in}_{\mp}(\Omega')\ket{\gamma}=
\left<e^{-K}\mathcal{O}^{in}_{\pm}(\Omega)e^Ke^{-K}
\mathcal{O}^{in}_{\mp}(\Omega')e^K\right>
 \ . \nonumber \\
\end{eqnarray}
In order to obtain these two point functions
we should calculate following expression
\begin{equation}
e^{-K}\mathcal{O}_{\pm}^{in}(\Omega)e^K=
\mathcal{O}_{\pm}^{in}(\Omega)+
\sum_{N=1}^{\infty}\frac{(-1)^N}{N!}
\overbrace{[K,[K,\mathcal{O}^{in}_{\pm}
(\Omega)],\dots,]]}^N \ .
\end{equation}
For this calculation we will need following
commutators
\begin{equation}
[\mathcal{O}^{out}_{\pm}(\Omega),\mathcal{O}^{in}
_{\mp}(\Omega')]=
-\mu\sinh\pi\mu\int d\Omega''
\triangle_{\mp}(\Omega,\Omega'')
[\mathcal{O}^{in}_{\mp}(\Omega''),
\mathcal{O}^{in}_{\mp}(\Omega')]=0
\end{equation}
and 
\begin{eqnarray}
[\mathcal{O}^{out}_{+}(\Omega)
,\mathcal{O}^{in}
_{+}(\Omega')]=
-\mu\sinh\pi\mu\int d\Omega''
\triangle_{-}(\Omega,\Omega'')
[\mathcal{O}^{in}_{-}(\Omega''),
\mathcal{O}^{in}_{+}(\Omega')]=\nonumber \\
=
-\mu\sinh\pi\mu\int d\Omega''
\triangle_{-}(\Omega,\Omega'')\frac{\mu}{2}
\delta (\Omega'',\Omega)
=-\frac{\mu^2}{2}\sinh\pi\mu \triangle_{-}(\Omega,\Omega')
\ , \nonumber \\
\left[\mathcal{O}^{out}_{-}(\Omega),\mathcal{O}^{in}
_{-}(\Omega')\right]=
-\mu\sinh\pi\mu\int d\Omega''
\triangle_{+}(\Omega,\Omega'')
[\mathcal{O}^{in}_{+}(\Omega''),
\mathcal{O}^{in}_{-}(\Omega')]=\nonumber \\
=
\mu\sinh\pi\mu\int d\Omega''
\triangle_{+}(\Omega,\Omega'')\frac{\mu}{2}
\delta (\Omega'',\Omega')
=\frac{\mu^2}{2}\sinh\pi\mu \triangle_{+}(\Omega,\Omega') 
\nonumber \\
\end{eqnarray}
so  we easily get
\begin{equation}
[K,\mathcal{O}^{in}_+(\Omega)]
=-2\overline{c}\mathcal{O}_+^{out}(\Omega)
 \ , 
\left[K,[K,\mathcal{O}_+^{in}(\Omega)]\right]=
4c\overline{c}\mathcal{O}_+^{in}(\Omega) \ .
\end{equation}
Then it is easy to see that
\begin{equation}
e^{-K}\mathcal{O}^{in}_+(\Omega)
e^K=N_{\gamma}(
\mathcal{O}^{in}_+(\Omega)-
e^{\gamma^*}
\mathcal{O}^{out}_+(\Omega)) \ 
\end{equation}
and consequently
\begin{eqnarray}
\bra{\gamma} \mathcal{O}^{in}_+(\Omega)
\mathcal{O}^{in}_+(\Omega')\ket{\gamma}=
\left<e^{-K}\mathcal{O}^{in}_+(\Omega)
e^K e^{-K}\mathcal{O}^{in}_+(\Omega'))
e^K\right>=\nonumber  \\
=N_{\gamma}^2\left<
(\mathcal{O}^{in}_+(\Omega)-
e^{\gamma^*}
\mathcal{O}^{out}_+(\Omega))
(\mathcal{O}^{in}_+(\Omega')-
e^{\gamma^*}
\mathcal{O}^{out}_+(\Omega')
\right>=\nonumber \\
=-N^2_{\gamma}e^{\gamma^*}
\left<\mathcal{O}^{out}_+(\Omega)
\mathcal{O}^{in}_+(\Omega')\right>=
N^2_{\gamma}e^{\gamma^*}\mu\sinh\pi\mu
\int d\Omega''\triangle_-(\Omega,
\Omega'')\left<\mathcal{O}^{in}_-(\Omega'')
\mathcal{O}^{in}_+(\Omega')\right>=\nonumber \\
=N^2_{\gamma}e^{\gamma^*}\mu\sinh\pi\mu
\int d\Omega''\triangle_-(\Omega,
\Omega'')\frac{\mu^2}{2}\delta(\Omega'',
\Omega)=\frac{\mu^2}{2}N^2_{\gamma}e^{\gamma^*}
\sinh\pi\mu
\triangle_-(\Omega,
\Omega') \ . \nonumber \\
\end{eqnarray}
Performing the same
analysis for  $\mathcal{O}^{in}_-$ 
leads to 
\begin{equation}
e^{-K}\mathcal{O}^{in}_-(\Omega)
e^K=
N_{\gamma}(\mathcal{O}^{in}_-(\Omega)
-e^{\gamma}\mathcal{O}^{out}_-(\Omega))
\end{equation}
and hence
\begin{eqnarray}
\bra{\gamma} \mathcal{O}^{in}_-(\Omega)
\mathcal{O}^{in}_-(\Omega')\ket{\gamma}=
\left<e^{-K}\mathcal{O}^{in}_-(\Omega)
e^K e^{-K}\mathcal{O}^{in}_-(\Omega')
e^K\right>=\nonumber  \\
=N_{\gamma}^2\left<
(\mathcal{O}^{in}_-(\Omega)-
e^{\gamma}
\mathcal{O}^{out}_-(\Omega))
(\mathcal{O}^{in}_-(\Omega')-
e^{\gamma}
\mathcal{O}^{out}_-(\Omega')
)\right>=\nonumber \\
=-N^2_{\gamma}e^{\gamma}
\left<\mathcal{O}^{in}_-(\Omega)
\mathcal{O}^{out}_-(\Omega')\right>=
N^2_{\gamma}e^{\gamma}\mu\sinh\pi\mu
\int d\Omega''\triangle_+(\Omega,
\Omega'')\left<\mathcal{O}^{in}_-(\Omega'')
\mathcal{O}^{in}_+(\Omega'')\right>=\nonumber \\
=N^2_{\gamma}e^{\gamma}\mu\sinh\pi\mu
\int d\Omega''\triangle_+(\Omega,
\Omega'')\frac{\mu}{2}\delta(\Omega'',
\Omega)=N^2_{\gamma}e^{\gamma}\frac{\mu^2}{2}\sinh\pi\mu
\triangle_+(\Omega,
\Omega') \ .  \nonumber \\
\end{eqnarray}
Finally we will calculate the mixed correlators
\begin{equation}
\bra{\gamma}\mathcal{O}^{in}_-(\Omega)
\mathcal{O}^{in}_+(\Omega')\ket{\gamma}=
N^2_{\gamma}\left<\mathcal{O}^{in}_-(\Omega)
\mathcal{O}^{in}_+(\Omega')\right>=
N^2_{\gamma}\frac{\mu}{2}\delta(
\Omega,\Omega')
\end{equation}
and
\begin{eqnarray}
\bra{\gamma}\mathcal{O}^{in}_+(\Omega)
\mathcal{O}^{in}_-(\Omega')\ket{\gamma}=
N^2_{\gamma}e^{\gamma+\gamma^*}
\left<\mathcal{O}^{out}_+(\Omega)
\mathcal{O}^{out}_-(\Omega')\right>=
\nonumber \\
=N^2_{\gamma}e^{\gamma+\gamma^*}
\mu^2\sinh^2\pi\mu\int
d\Omega''d\Omega'''\triangle_-(\Omega,\Omega'')
\frac{\mu}{2}\delta(\Omega'',\Omega''')
\triangle_+(\Omega,\Omega''')=
\frac{\mu}{2}
N^2_{\gamma}e^{\gamma+\gamma^*}
\delta(\Omega,\Omega') \ .   \nonumber \\
\end{eqnarray} 
These two point functions  agree with the correlators that  were
calculated 
in \cite{Bousso:2001mw,Spradlin:2001nb} from
slightly different approach. We have seen that
these correlators can be easily calculated 
using the conjecture that the states 
in  the dual CFT  correspond to the  states
of massive scalar field in   de Sitter space
at $\mathcal{I}^-$.  
\section{Conclusion}\label{fifth}
In this paper we 
 tried to present our point of view
on the nature of dS/CFT correspondence.
We studied the relation between
free massive scalar  field in the bulk of 
de Sitter space and the  dual Euclidean
CFT living in the past infinity
$\mathcal{I}^-$. 
Using the  definition of the inner product in
the Hilbert space
$\mathcal{H}_i$ of initial states
of massive scalar field in de Sitter space
 we were able to reconstruct
all two point functions of dual CFT operators in arbitrary
vacuum state. 
Let us express  once again the main idea of
our approach. Following arguments given
in \cite{Witten:2001kn} we consider 
operators in  dual CFT 
living on $\mathcal{I}^-$ that
define states 
in  the Hilbert space $\mathcal{H}_i$ 
of  initial states of quantum gravity in 
de Sitter space. Using this conjecture we
can from the  known  inner product and
the structure of the Hilbert space of initial states
on $\mathcal{I}^-$  obtain all 
correlators in the dual CFT and define adjoint
operation. We have seen that  application
of this idea gives   correlators in 
the dual CFT that have the same form as the
two point functions that  were calculated in 
\cite{Bousso:2001mw,Spradlin:2001nb} 
from  the equivalence of
 partition function of CFT on $\mathcal{I}^-$ 
 and S-matrix elements  in the bulk. 

It is clear that the same analysis,
as was presented in previous sections,   could be
performed for CFT defined on $\mathcal{I}^+$ in
the sense that this theory gives an exact description
of the Hilbert space $H_f$ of the final states
of massive scalar field  at past infinity. Of course in
the case of the  example  studied in
this paper these two Hilbert spaces 
are identical  and hence corresponding
CFT are the same. For that reason  we considered
the  CFT defined on $\mathcal{I}^-$ as the fundamental one.
However it is clear that one single CFT cannot
contain information about dynamical evolution of
the bulk quantum gravity which, in our personal
point of view, should have its CFT description 
in terms of renormalization group flow 
\cite{Strominger:2001gp} that maps CFT operators
on $\mathcal{I}^+$ corresponding to UV limit of
the theory to the CFT operators on
$\mathcal{I}^-$ corresponding to IR limit.

The extension of this approach is natural. The first
possibility  is to apply this analysis to the case of 
arbitrary $\mu$ as in
\cite{Leblond:2002tf,Tolley:2001gg}.
It would be also nice to
analyse 
massless bulk scalar field with its potential  application to 
the cosmology
\cite{Danielsson:2002qh,Bergstrom:2002yd,Danielsson:2002kx,
Danielsson:2002mb}. Very interesting analysis
of this problem from the point of view of dS/CFT was
given recently in \cite{Maldacena:2002vr}.
We would like also study  the relation between
CFT and quantum field theory  in  
Elliptic de Sitter space  \cite{Parikh:2002py}
and the relation between CFT and
quantum field theory in  the space-times that 
approach their de Sitter phase in the asymptotic
past or future only.

Even if we know that this paper gives only modest
contribution to the question of the quantum gravity
in de Sitter space and its relation to the string theory
we   believe that this small contribution  could be helpful for
further study of  dS/CFT correspondence.
\\
\\
{\bf Acknowledgment}
We have benefitted greatly from  
discussions with  Ulf Danielsson.
This work is partly supported by EU contract
HPRN-CT-2000-00122.

\end{document}